\documentclass[12pt]{article}
\usepackage{amssymb,amsmath,cite}
\long\def\@makefntext#1{\parindent 0cm\noindent
\hbox to 1em{\hss$^{\@thefnmark}$}#1}

\newcommand{\beq}{\begin{equation}}
\newcommand{\eeq}{\end{equation}}

\begin{document}
\vspace{.5in}
\begin{flushright}
UCD-05-12\\
gr-qc/0510056\\
August 2005\\
\end{flushright}
\vspace{.5in}
\begin{center}
{\large\bf
 Reply to ``Comment on Model-dependence of Shapiro time delay\\[.8ex]
 and the `speed of gravity/speed of light' controversy''}\\
\vspace{.4in}
{S.~C{\sc arlip}\footnote{\it email: carlip@physics.ucdavis.edu}\\
       {\small\it Department of Physics}\\
       {\small\it University of California}\\
       {\small\it Davis, CA 95616, USA}}
\end{center}
\addtocounter{footnote}{-1}

To determine whether Shapiro time delay---or any other measured
quantity---depends on the ``speed of gravity'' $c_g$ or the 
``speed of light'' $c$, one must analyze observations in
a framework in which $c_g$ and $c$ can be different.  A model
with $c_g\ne c$ is almost inevitably a bimetric theory, with
a ``gravitational metric'' $g_{ab}$ whose null cones describe
propagation at speed $c_g$ and an ``electromagnetic metric'' 
${\tilde g}_{ab}$ whose null cones describe propagation at 
speed $c$.  In \cite{Carlip}, I investigated the observation 
by Fomalont and Kopeikin \cite{FomKop} of the Shapiro time 
delay of quasar light passing near Jupiter, using a particular 
bimetric model studied by Jacobson et al.\ \cite{Jacobson}.  I 
concluded that, contrary to some claims in the literature, the 
leading velocity-dependent term in the delay depended on $c$ 
and not on $c_g$.

In his comment \cite{Kopeikin}, Kopeikin argues that my results 
have an additional, hidden dependence on $c_g$, coming from the 
choice of units for the spatial coordinate ${\bf x}$.  More 
precisely, the expression (4.13) in \cite{Carlip} for the time 
delay involves certain physical quantities, the ``instantaneous''
relative separation ${\bf r}$ of of Jupiter and the Earth and 
the velocity ${\bf v}_J$ of Jupiter.  Kopeikin argues that these 
are implicitly given in ``gravitodynamic'' units, which differ 
from SI units by factors of $c/c_g$.

This is an important and subtle issue, and I expect that it is
relevant for some bimetric models.  In the particular model I 
have used, though, I believe Kopeikin's conclusion is not correct.  
The issue is not simply that I chose to set $c_g$ to unity; factors 
of $c_g$ may be restored trivially by simple dimensional analysis.  
Nor can the physical significance of a coordinate such as ${\bf r}$
be simply read off from one of the two metrics in a bimetric 
theory.  Rather, one must understand ${\bf r}$ and ${\bf v}_J$ as 
\emph{observables}, that is, one must understand how the model 
tells us they are measured.  To do so, note the following features 
of the model used in \cite{Carlip}:
\begin{enumerate}
\item Any clock or other measuring instrument made from ordinary
matter and electromagnetic fields has an action that depends solely
on the electromagnetic metric ${\tilde g}_{ab}$ (see \cite{Carlip},
eqn.\ (2.5)).  As a result, nongravitational measurements of time
and distance will depend on $c$ and not $c_g$, and can be taken to
be ordinary ``SI'' measurements.  In a gravitational field, there 
may be corrections involving $c_g$ coming from $c_g$ dependence
of gravitational time dilation, but these will be extremely small.
\item Light propagates along null geodesics of the electromagnetic 
metric.  In particular, in the flat space limit, the time for light
to travel from ${\bf x_e}$ to ${\bf x}$ is
\beq
t-t_e = \frac{|{\bf x} - {\bf x_e}|}{c}
\eeq
(see \cite{Carlip}, eqn.\ (4.8)).  As a result, distances measured
by light propagation again depend, to first order, on $c$ and not 
$c_g$.
\item The motion of a point mass---say, Jupiter---is given by a 
timelike geodesic in the electromagnetic metric (see \cite{Carlip}, 
eqn.\ (2.7)).  As long as the coupling is properly normalized (see
\cite{Carlip}, eqn.\ (3.18)), positions and velocities inferred 
from fits to a Newtonian or post-Newtonian model will agree with 
standard ``SI'' results.
\end{enumerate}
Thus, as \emph{observables}, ${\bf r}$ and ${\bf v}_J$ have no
hidden dependence on $c_g$.  The distance ${\bf r}$, for example, 
can be measured within the model by using an atomic clock (with 
dynamics depending only on $c$) to time a light pulse traveling 
from Earth to Jupiter at speed $c$. 

The absence of any hidden dependence on $c_g$ in \cite{Carlip} may 
be further checked by restoring all factors of $c_g$ and making all 
units explicit.  The relevant term in the Shapiro time delay comes 
from an integral of the Newtonian potential, $\int_{t_e}^0\phi\, dt$,
along the light path, where I have set the arrival time to $t=0$.
As Kopeikin agrees (see \cite{Kopeikin}, eqn.\ (12)), the Newtonian
potential $\phi$ is
\beq
\phi(t,{\bf x}) = \frac{GM_J}{|{\bf x}-{\bf x}_J(t)|} 
  + {\cal O}\left(\frac{v_J^2}{c_g^2}\right) .
\label{a1}
\eeq
The unperturbed light path is a straight line, and to first order in
velocity, so is Jupiter's path.  Expanding to this order, we have
\beq
{\bf x}-{\bf x}_J(t) 
  \approx {\bf x}(0) + {\bf k}t - {\bf x}_J(0) - {\bf v}_Jt
  = {\bf r} + ({\bf k} - {\bf v}_J)t .
\eeq
The integral of $\phi$ can now be performed exactly; one obtains
\beq
\int_{t_e}^0\phi\,dt = \frac{GM_J}{|{\bf k} - {\bf v}_J|} \ln\left[
  \frac{1}{2r_*}\left(
  r - \frac{({\bf k} - {\bf v}_J)\cdot{\bf r}}{|{\bf k} - {\bf v}_J|}
  \right)\right]
\label{a2}
\eeq
where $r_*$ is an (irrelevant) distance to an initial fiducial point
(see \cite{Will} for details).  The argument of the logarithm is
dimensionless, and thus independent of any consistently chosen system 
of units.  But now note that in the model discussed in \cite{Carlip},
\beq
|{\bf k}|^2 = c^2 .
\eeq
An elementary calculation then shows that to first order in $v_J$,
\beq
\frac{({\bf k} - {\bf v}_J)\cdot{\bf r}}{|{\bf k} - {\bf v}_J|}
  = \left({\widehat{\bf k}} - {\boldsymbol\beta} + {\widehat{\bf k}}\cdot
  {\boldsymbol\beta}\,{\widehat{\bf k}}\right)\cdot{\bf r}
  = \left( {\widehat{\bf k}} - {\widehat{\bf k}}\times({\boldsymbol\beta}
  \times{\widehat{\bf k}})\right)\cdot{\bf r}
\eeq
with ${\boldsymbol\beta} = {\bf v}_J/c$, in agreement with \cite{Carlip}.  
We thus see that the dependence on $c$ in the argument of the logarithm
(\ref{a2}) comes directly from the fact that $|{\bf k}|^2 = c^2$, that 
is, from the fact that light travels at speed $c$.

Let me conclude with a few words about the meaning of this result.
Intuitive interpretations are always a bit delicate in a generally 
covariant theory, since one's intuition often involves hidden assumptions 
about coordinates, but I believe the following picture may be helpful.  
As Kopeikin points out at the end of \cite{Kopeikin}, velocity dependence 
arises in three places in this computation: in the retardation of the 
gravitational field, the aberration of gravity, and the aberration of 
light.  Eqn.\ (\ref{a1}) describes the approximate cancellation of the 
first two of these, leading to an apparently ``instantaneous'' potential. 
As discussed in \cite{Carlipb}, this cancellation can be obtained from 
conservation of energy in any theory in which gravitational radiation 
couples only to quadrupole and higher moments of the stress-energy tensor.  
Since this cancellation can be deduced from a system in which the relevant 
interactions are purely gravitational---a binary pulsar, for example---it 
must involve the two terms dependent on $c_g$.  This leaves the aberration 
of light, which depends on $c$.  It should therefore come as no surprise 
that the final time delay depends on $c$ rather than $c_g$.

These considerations may also point to another class of models in which
the time delay might depend on $c_g$ as well: bimetric theories that
permit monopole or dipole gravitational radiation, thus negating the
need for a cancellation between gravitational retardation and aberration.
Many such models may be ruled out by other observations, but a complete
analysis is lacking, and the observations of \cite{FomKop} may further
limit the parameter space.

\vspace{1.5ex}
\begin{flushleft}
\large\bf Acknowledgments
\end{flushleft}

\noindent This work was supported in part by Department of Energy grant
DE-FG02-91ER40674.


\begin{thebibliography}{99}
\bibitem{Carlip} S.\ Carlip, Class.\ Quant.\ Grav.\ 21 (2004) 3803, 
 gr-qc/0403060.
\bibitem{FomKop} E.\ B.\ Fomalont and S.\ M.\ Kopeikin, Astrophys.\ J.\ 598
 (2003) 704, astro-ph/0302294.
\bibitem{Jacobson} T.\ Jacobson and D.\ Mattingly, Phys.\ Rev.\ D64 (2001) 
 024028.
\bibitem{Kopeikin} S.\ M.\ Kopeikin, gr-qc/0510048.
\bibitem{Will} C.\ M.\ Will, Astrophys.\ J.\ 590 (2003) 683, astro-ph/0301145.
\bibitem{Carlipb} S.\ Carlip, Phys.\ Lett.\ A267 (2000) 81, gr-qc/9909087.

\end{thebibliography}
\end{document}